\documentclass[aps,pra,twocolumn,superscriptaddress]{revtex4-1}

\usepackage{graphicx}
\usepackage{hyperref}
\usepackage{amsmath}
\usepackage{amssymb}
\usepackage{epstopdf}
\usepackage{multirow}
\usepackage{threeparttable}
\usepackage[usenames]{color}

\usepackage{graphicx}
\begin{document}

\title{Infrared Single-Pixel Hyperspectral Imaging via Spatial-Temporal Multiplexing}

\author{Ben Sun}
\affiliation{State Key Laboratory of Precision Spectroscopy, and Hainan Institute, East China Normal University, Shanghai 200062, China}

\author{Kun Huang}
\email{khuang@lps.ecnu.edu.cn}
\affiliation{State Key Laboratory of Precision Spectroscopy, and Hainan Institute, East China Normal University, Shanghai 200062, China}
\affiliation{Chongqing Key Laboratory of Precision Optics, Chongqing Institute of East China Normal University, Chongqing 401121, China}
\affiliation{Collaborative Innovation Center of Extreme Optics, Shanxi University, Taiyuan, Shanxi 030006, China}

\author{Zhibin Zhao}
\affiliation{State Key Laboratory of Precision Spectroscopy, and Hainan Institute, East China Normal University, Shanghai 200062, China}

\author{Beibei Dong}
\affiliation{State Key Laboratory of Precision Spectroscopy, and Hainan Institute, East China Normal University, Shanghai 200062, China}

\author{Jianan Fang}
\affiliation{State Key Laboratory of Precision Spectroscopy, and Hainan Institute, East China Normal University, Shanghai 200062, China}
\affiliation{Chongqing Key Laboratory of Precision Optics, Chongqing Institute of East China Normal University, Chongqing 401121, China}

\author{Heping Zeng}
\email{hpzeng@phy.ecnu.edu.cn}
\affiliation{State Key Laboratory of Precision Spectroscopy, and Hainan Institute, East China Normal University, Shanghai 200062, China}
\affiliation{Chongqing Key Laboratory of Precision Optics, Chongqing Institute of East China Normal University, Chongqing 401121, China}
\affiliation{Shanghai Research Center for Quantum Sciences, Shanghai 201315, China}
\affiliation{Chongqing Institute for Brain and Intelligence, Guangyang Bay Laboratory, Chongqing, 400064, China}

\begin{abstract}
Near-infrared (NIR) hyperspectral imaging is widely used to reveal morphological and chemical information. However, conventional spectral imagers usually rely on costly focal plane arrays and suffer from data redundancy and inefficiencies in spatial-spectral data acquisition. Here, we devise and implement a single-pixel NIR hyperspectral imaging system based on high-fidelity spectrum-to-time mapping and high-precision spatial-encoding compressive measurements. The system employs a single-mode telecommunication fiber for temporal dispersion and a programmable spatial light modulator to impose structured spatial patterns, with all signals detected by a single InGaAs photodetector. By correlating temporally stretched waveforms with spatial encodings, we reconstruct 64$\times$64 spatially resolved hyperspectral datacubes spanning 50 spectral bands over the 1550-1600 nm range. Furthermore, real-time monitoring of dynamic liquid injection is demonstrated at a datacube refreshing rate of 12 Hz under sub-Nyquist sampling. The presented architecture features single-pixel simplicity, high optical throughput, and efficient data acquisition, which would pave a novel way for NIR spectral imaging in biomedical diagnostics and material characterization.
\end{abstract}

\maketitle

\section{Introduction}
Near-infrared (NIR) hyperspectral imaging has emerged as an indispensable tool for resolving chemical constituents and distribution by simultaneously capturing spatial-spectral information \cite{Khan2018IEEE}. Benefiting from the characteristic overtones and combination bands of fundamental vibrational modes occurring in the NIR region, NIR spectroscopy provides a rich source of chemical information about material compositions and molecular interactions \cite{Ozaki2021Book}. Currently, NIR hyperspectral imaging demonstrates broad applicability in the fields of biology, medicine, and materials science, such as non-destructive pathological diagnosis, rapid industrial detection, label-free environmental monitoring, and non-invasive agricultural surveillance \cite{Wang2024NP, Rodrigues2022ABC, Stegemann2025AS, Liu2017TFST}. In these scenarios, spectral analysis is often conducted through spatial dispersion devices \cite{Li2024NC, Arce2014ISPM} or spectral filtering configurations \cite{Roxbury2015SR, Kaariainen2021OL}, which introduce system complexity and redundancy. Furthermore, the mandatory use of NIR focal plane arrays (FPAs) imposes additional costs on hyperspectral imaging systems \cite{Li2018IPT}. In addition, Fourier-transform infrared (FTIR) spectroscopic imaging has evolved over decades in chemical imaging \cite{Dorling2013TB}, with the majority of modern systems predominantly operating through FPAs to enhance data acquisition speed. Notably, spectral phasor transformation technology has demonstrated significant promise in NIR hyperspectral imaging \cite{Stegemann2025AS}, although its implementation still necessitates costly FPA for 2D spectral (phasor) space acquisition. Recent advances in on-chip computational hyperspectral imagers have emerged as promising candidates for 3D spectral data cube reconstruction \cite{Wang2019NC, Bian2024Nature, Tao2021OE}. Nevertheless, the complexities in fabrication processes and the stability against external disturbances persist as critical challenges for their deployment in specialized scenarios. Therefore, it is imperative to develop novel technologies enabling highly efficient and adaptive NIR hyperspectral imaging.

In this context, single-pixel imaging has emerged as an innovative imaging architecture to circumvent the aforementioned limitations for high-definition FPAs \cite{Edgar2019NP}. Notably, this technique has been explored in diverse domains including spectral, holographic, and 3D imaging \cite{Zhang2018Optica, Wu2021NC, Sun2016NC}, where its single-element detection paradigm provides a cost-effective alternative to multi-pixel counterparts. Recently, the integration of a single-pixel camera with spectroscopic techniques has demonstrated promising potential in spectral imaging research. Current implementations of single-pixel spectral imagers primarily achieve spectral analysis through spectrographs \cite{Pian2017NP, Uguen2024APL} or spectral filtering devices \cite{Welsh2013OE, Gattinger2019OE}. The former approach typically requires a multi-pixel detector, which is inherently limited in acquisition rate for dynamic scenes. Moreover, when scaling to larger numbers of spectral channels, the need for high-definition pixel formats becomes both technically demanding and economically prohibitive. Furthermore, the interdependence between pixel count and number of spectral channels imposes a fundamental trade-off between spectral bandwidth and spectral resolution. The latter methodology demands high-performance filters to enhance spectral resolution. Both single-pixel spectral imagers inevitably constrain optical throughput, leading to diminished photon utilization efficiency. Nowadays, the use of spatial light modulators (SLMs) offers a viable approach for spectral differentiation by leveraging their programmable optical manipulation \cite{Ebner2023SR, Xu2024NC, Sun2025LPR}. Recent attempts employing spectral modulation module \cite{Bian2016SR} or interferometer \cite{Jin2017SR} have demonstrated the feasibility of single-pixel spectral imaging through multiplexing spatial-spectral information to the temporal sequence, though their dependence on mechanical parts introduces inertial errors and vibration-induced sensitivity. Moreover, emerging materials with distinct spectral responses, such as quantum dots \cite{Meng2024LSA} and metasurfaces \cite{Xiong2023LSA}, have been applied for single-pixel spectral imaging. However, these architectures have limitations in expanding the number of spectral channels. To date, it remains appealing to acquire rapid, high-resolution, and high-throughput spectral information for single-pixel hyperspectral imaging technology.

Recently, photonic time-stretch has been established as a rapid and efficient spectroscopy approach, also known as dispersive Fourier transform (DFT) \cite{Solli2008NP, Goda2013NP, Hashimoto2023LSA}. DFT maps the spectrum of a pulse to a temporal waveform through a dispersive medium with large group velocity dispersion (GVD), allowing a single-pixel detector to measure spectral information matching the temporal intensity profile \cite{Goda2013NP, Sun2024LPR}. Particularly, the unique spectrum-to-time mapping has been inspired to establish the so-called optical time-stretch imaging paradigm \cite{Zhang2025NRMP, Cheng2016APR}. By exploiting spatio-temporal dispersion properties, spatial information of the target can be spectrally encoded by establishing distinct frequency-to-space correspondence \cite{Goda2009Nature, Lau2016LOAC}. Additionally, temporally structured illumination generated by encoding time-stretched signals has been investigated in 3D imaging \cite{Teng2020AP} and microscopy \cite{Duan2019OL}. However, current frequency-to-space mapping implementations cannot simultaneously capture full-field images per spectral channel due to the spatial information inherently occupying the spectral dimension. A scheme based on spectrum-time mapping and compressive imaging has been reported to probe the whole spatial-spectral distribution of a pulsed laser illuminating a diffuser \cite{Thomas2018OE}, yet the full potential of its hyperspectral imaging power has not been realized. So far, it remains a challenging endeavor to advance time-stretch spectroscopy to the spectral imaging field.

In this work, we have devised and implemented a single-pixel NIR hyperspectral imaging system, leveraging both merits of time-stretch spectroscopy and single-pixel imaging. Specifically, a section of single-mode fiber is used as the dispersive medium to impose temporal dispersion on NIR ultrashort pulses for establishing a spectrum-to-time mapped optical source. Subsequently, a programmable digital micromirror device (DMD) is used to perform the high-fidelity spatial encoding. The spatiotemporally modulated signal is then registered by a high-bandwidth single-element InGaAs detector. Finally, the hyperspectral image is reconstructed with 50 spectral bands across 1550-1600 nm with a pixel resolution of 64$\times$64. Moreover, a real-time monitoring of liquid injection dynamics at a refreshing rate of 12 Hz is demonstrated by combining the compressive sensing algorithm. Our approach extends traditional single-pixel imaging to 3D hyperspectral imaging through wavelength-to-time mapping, where temporally encoded signals are embedded within pattern-switching intervals without introducing additional temporal overhead. The presented system enables high-throughput and high-speed hyperspectral imaging with single-pixel detection, offering an accessible and cost-effective solution for biomedical diagnostics and materials characterization.

\begin{figure*}[t!]
\includegraphics[width=0.85\textwidth]{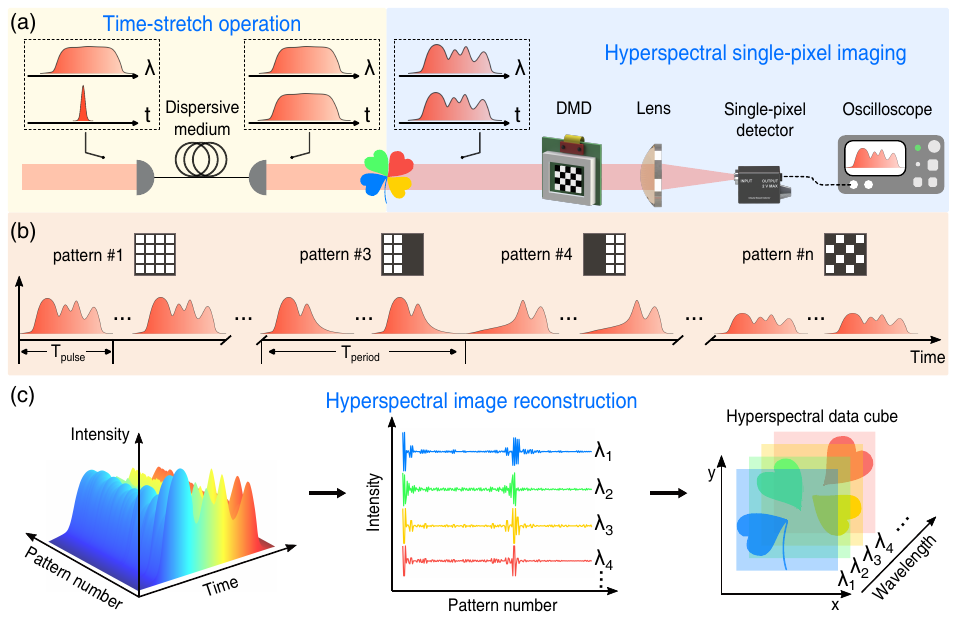}
\caption{Conceptual illustration of the single-pixel time-stretch NIR hyperspectral imaging. (a) Schematic diagram for the hyperspectral data acquisition. A broadband NIR source is sent into a dispersive medium for performing the time-stretch operation. The temporally stretched pulse then illuminates a sample to carry spatial and spectral information. A digital micromirror device (DMD) is used to facilitate the spatial encoding. Consequently, the encoded signal is detected by a single-pixel photodiode and recorded by an oscilloscope. (b) Measured time-domain sequences under different illumination patterns. (c) Reconstruction process of the hyperspectral image. The wavelength-to-time mapping enables the extraction of the intensity values associated with encoding patterns for each spectral band. The spectral data cube can be reconstructed by combining the knowledge of the projected patterns and the corresponding measured intensities.}
\label{fig1}
\end{figure*}
 
\section{Basic principle}
The conceptual illustration of the proposed system for hyperspectral imaging is illustrated in Fig. \ref{fig1}, comprising three core parts: time-stretch operation, single-pixel spectral imaging, and hyperspectral image reconstruction. The generation of a time-stretched pulsed optical source serves as a critical prerequisite for system operation. The implementation of time-stretch modulation is achieved by the propagation of optical pulses through a dispersive medium, governed by the nonlinear Schr$\ddot{\text{o}}$dinger equation with linear gain and loss terms, as mathematically expressed by \cite{Goda2013NP}:
\begin{equation}
|E(L,T)|^2 =  \frac{2}{\pi \beta L} e^{-\alpha L} |\tilde{E}(0, \frac{T}{\beta L})|^2 \ 
\label{eq1}
\end{equation}
where $E$ is the field amplitude of the optical pulse, $\tilde{E}$ is the Fourier transform of the signal, $\alpha$ is the absorption coefficient, $L$ is the propagation distance, $\beta$ is the second-order dispersion coefficient, and $T$ is the relative time in the moving frame of the propagation pulse. The resulting one-to-one transformation precisely maps the wavelength of the optical pulse to a temporal waveform according to $\delta t = D L \delta \lambda$, where $\delta t$ is the time-stretched duration, $D$ is the GVD introduced by the dispersive medium per unit length, and $\delta \lambda$ is the spectral bandwidth of the optical pulse. The underlying mechanism leverages the space-time duality between far-field spatial diffraction and dispersive pulse propagation in the time domain \cite{Zhang2025NRMP}. Distinct from conventional spectral analysis methods, the time-stretch operation eliminates requirements for spatial dispersive elements, narrowband filters, or mechanical moving parts.

To extend the time-stretch spectroscopy to hyperspectral imaging, we integrate a SLM to impose time-varying spatial encoding as shown in Fig. \ref{fig1}(a). This configuration enables multiplexed detection of spatiotemporal signals via a single-pixel detector, thus circumventing the reliance on costly FPAs. Note that the use of high-speed single-pixel detectors allows one to capture additional information during the elapse of each spatially modulated pattern of the DMD without introducing additional acquisition time \cite{Bian2016SR, Jiang2022PR}.

\begin{figure*}[t!]
\includegraphics[width=0.8\textwidth]{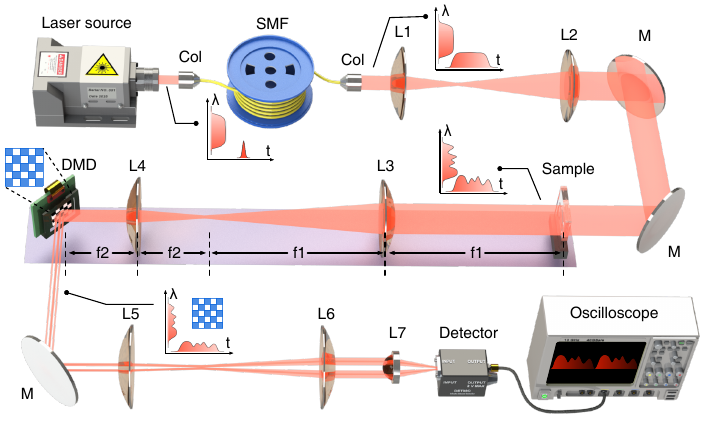}
\caption{Experimental setup. A NIR laser source is used to deliver broadband ultrafast pulses. The signal pulse is coupled into a long haul of single-mode fiber (SMF) to perform the time-stretch operation. The resultant time-stretch pulse is collimated back to the free space, and the beam size is expanded by two lenses (L1 and L2). Then, the signal beam illuminates a sample in a 4f imaging system formed by L3 and L4, where the DMD is placed at the image plane. The modulated signal is steered into another 4f system (L5 and L6) to match the aperture size of the focusing lens (L7). Finally, the encoded time-domain signal is registered by a single-pixel detector, and acquired via a digital oscilloscope.. Correlating the recorded temporal waveforms with the associated projected patterns allows us to reconstruct the hyperspectral image.}
\label{fig2}
\end{figure*}

A hyperspectral image reconstruction algorithm can be readily derived from the single-pixel imaging modality. Given that the repetition rate ($>$MHz) of ultrafast pulse significantly exceeds the frame switching rate ($\sim$kHz) of the SLM, each illumination period $T_\text{period}$ of the time-varying projected pattern contains thousands of identical time-stretched waveforms as illustrated in Fig. \ref{fig1}(b). This process ultimately generates encoded temporal waveforms corresponding to a sequence of well-defined patterns. Specifically, we vectorize the hyperspectral image into a 2D matrix $ x^{t, \lambda} \in \mathbb{R}^{N \ast L} $, where $N = N_x \times N_y$ denotes the total number of spatial pixels and $L$ is the number of spectral channels. Each column of $ x^{t, \lambda}$ corresponds to the spatial intensity distribution at different spectral channels. According to the calibrated wavelength-to-time mapping relation, the temporal axis $t$ can be directly converted to the wavelength axis $\lambda$. The measurement process can be represented as:
\begin{equation}
y^{t, \lambda} = \Phi x^{t, \lambda},
\label{eq2}
\end{equation}
where $ y^{t, \lambda} \in \mathbb{R}^{N \ast L} $ is the measured intensity at different spectral channels and $ \Phi \in \mathbb{R}^{N \ast N} $ is the measurement matrix. The hyperspectral data cube can be reconstructed by solving the inverse problem as shown in Fig. \ref{fig1}(c). Compared with 2D imaging, 3D hyperspectral imaging exhibits higher intrinsic redundancy and enhanced compressibility. The application of compressive sensing enables high-quality reconstruction of redundant signals at sampling rates far below the Nyquist-Shannon requirement \cite{Duarte2008IEEE}. The compressive sensing operation using M (M $\textless$ N) measurements can be expressed as:
\begin{equation}
y^{t, \lambda} =  \Phi x^{t, \lambda} = \Phi \Psi s^{t, \lambda} = \Theta s^{t, \lambda},
\label{eq3}
\end{equation}
where $y^{t, \lambda} \in \mathbb{R}^{M \ast L}$, $ \Phi \in \mathbb{R}^{M \ast N}$, $\Theta = \Phi \Psi \in \mathbb{R}^{M \ast N}$, $s^{t, \lambda} \in \mathbb{R}^{N \ast L}$ is a collection of sparse coefficients of $x^{t, \lambda}$ in basis $\Psi$. To achieve this, we utilized the Walsh-ordered Hadamard matrix as the measurement matrix, ensuring incoherence with the sparse basis. The total variation augmented Lagrangian (TVAL3) algorithm is implemented for efficient hyperspectral image reconstruction. More details about the compressive reconstruction algorithm can be found in our previous work \cite{Sun2025LPR}.

\begin{figure*}[t!]
\includegraphics[width=0.8\textwidth]{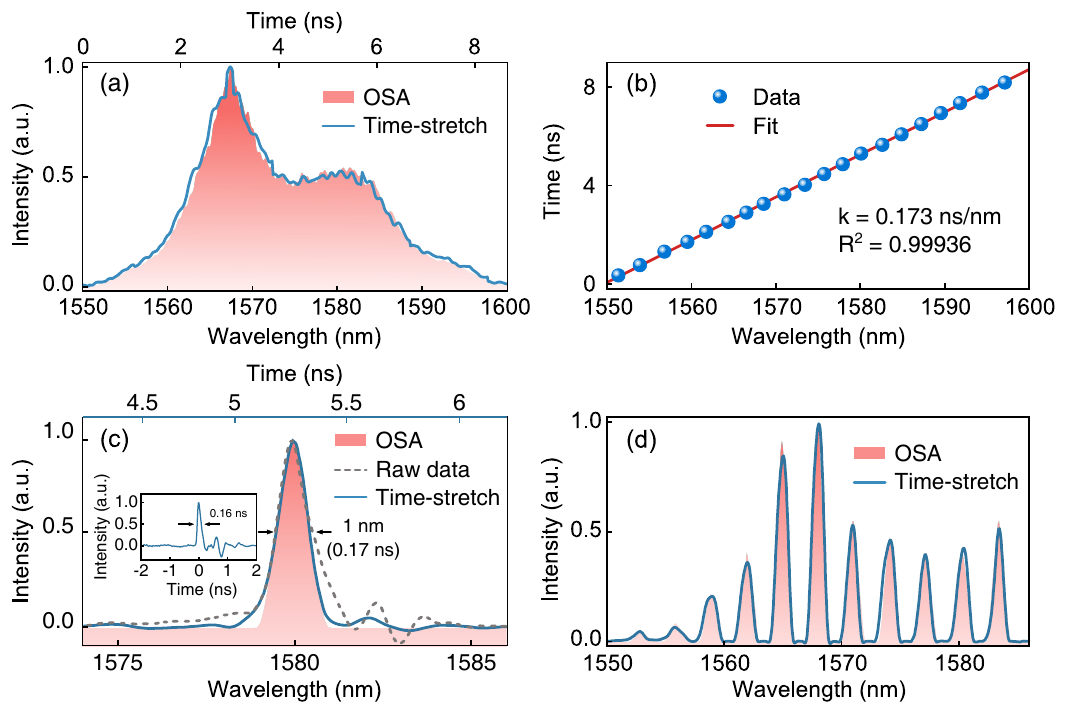}
\caption{Spectro-temporal characterization of the time-stretched optical pulses. (a) Recorded optical spectrum (shaded area) for the NIR signal by an optical spectrum analyzer (OSA), and the blue line indicates the time-stretched waveform. (b) Measured time delay as a function of the center wavelength by tuning the programmable optical filter with a 1-nm spectral resolution. The solid line denotes a linear fit. (c) A representative illustration of the filtered spectrum (shaded area) by the OSA, the raw data (dashed line) of the measured temporal waveform, and the corrected waveform (blue line) after the deconvolution operation for the detector's impulse response (inset) (d) Measured optical spectrum in the case of multi-wavelength filtering.}
\label{fig3}
\end{figure*}

\section{Experimental results}
Figure \ref{fig2} illustrates the experimental setup for the time-stretch NIR hyperspectral imaging with single-pixel detection. The involved light source originated from an Er-doped mode-locked fiber laser (EDFL, LangyanTech, ErFemto Elite) operating at a repetition rate of 82.1 MHz. The EDFL emitted ultrafast pulses with a temporal duration of 260 fs, providing a stable and broadband laser source. Then, the NIR femtosecond pulses were coupled into a single-mode fiber (SMF, YOFC, G652D) with a length of 10 km to perform the time-stretch operation. The SMF exhibits a propagation loss around 0.22 dB km$^{-1}$ and a GVD about 17.1 ps (nm·km)$^{-1}$ at 1550 nm. The pulse output after fiber propagation maintained an average power of about 10 mW. Notably, Raman amplification implemented directly within the dispersive medium would enable significant enhancement of the average optical power, thereby offsetting dispersion-induced attenuation limits and facilitating weak signal recovery during pulse propagation \cite{Solli2008NP}. These generated time-stretched pulses were essential for subsequent hyperspectral single-pixel imaging implementation.

Then, the NIR time-stretched beam was expanded to illuminate the sample before being steered into a 4f imaging system consisting of two lenses with focal lengths of 200 and 50 mm. In the present configuration, placing the time-stretch module before the sample is critical: it suppresses nonlinear effects on the sample, preserves coupling efficiency by avoiding spatial mode degradation, and enhances the simplicity and robustness of the system architecture \cite{Shoshin2025LPR}. A DMD (Texas Instruments DLP650LNIR) positioned at the image plane performs spatial encoding via time-varying patterns. The DMD supports binary pattern projection at a frame rate up to 10.752 kHz through high-speed micromirror switching, which can operate at 800-2000 nm spectral range while maintaining $\textgreater$60$\%$ diffraction efficiency. Subsequently, the encoded beam was expanded to match the aperture size of a molded aspheric lens (f=4.6 mm, LBTEK, MAC90653-C), before being focused onto a free-space InGaAs photodetector with a 5 GHz bandwidth (Thorlabs, DET08C). Finally, the encoded time-stretched sequence was recorded by an oscilloscope (Keysight, DSA91304A) with a 10-GHz bandwidth operating at a sampling rate of 40 GS s$^{-1}$. Here, the encoding process employed differential measurements, effectively eliminating background-induced offsets in the image. The suppression of low-frequency oscillations from the illumination source \cite{Edgar2019NP} further enhances the reconstruction quality of the hyperspectral image. Note that a balanced detection scheme employing two photodiodes to capture both reflection ports of the DMD can significantly improve signal collection efficiency. This eliminates the necessity of projecting complementary patterns, thereby reducing the total number of patterns required and effectively doubling the system's frame rate. Furthermore, the electronic timing and data acquisition processes were managed by a controlling unit based on a field programmable gate array, which ensured precise acquisition of encoded temporal waveforms. More details on the operational configuration and timing sequence can be found in our previous work \cite{Wang2023NC}.

We start to calibrate the spectro-temporal characterization of the single-pixel time-stretch NIR hyperspectral imager. Figure \ref{fig3}(a) presents high-fidelity wavelength-to-time mapping between the NIR spectrum and the temporal waveform, establishing the foundation for hyperspectral image reconstruction. The precise wavelength-to-time mapping relation was investigated by inserting a programmable optical filter (Waveshaper, 1000A) before the SMF. Based on liquid crystal on silicon technology, the programmable optical filter is capable of independent control and processing of different wavelength components. The device offers programmable filtering bandwidths ranging from 10 GHz to 5.36 THz with attenuation control up to 35 dB. Figure \ref{fig3}(b) presents the mapping relation between the time delay and the central wavelength by tuning the programmable optical filter with a 1-nm spectral resolution. A slope obtained from the linear fitting is 0.173 ns nm$^{-1}$, which corresponds to the total GVD. However, due to the imperfect impulse response of the detector, the raw temporal waveform exhibits a minor mismatch compared to the spectrum measured by the OSA as shown in Fig. \ref{fig3}(c). The spectral resolution of about 1 nm (4 cm$^{-1}$) was determined by the full width at half maximum of the system's impulse response, corresponding to a measured duration of 160 ps in the presence of unstretched femtosecond NlR pulses. To mitigate the mismatch introduced by the system's impulse response, we apply the deconvolution method, followed by smoothing operation with a moving window on the raw temporal waveform \cite{Cai2024SA}. Figure \ref{fig3}(d) presents multi-wavelength filtering results after the correction, indicating a good agreement between the time-stretch spectrum and the OSA reference. The programmable optical filter was configured with a 1 nm spectral resolution per channel, specified with equal intervals. Note that the filter was removed after the calibration.

\begin{figure*}[t!]
\includegraphics[width=0.85\textwidth]{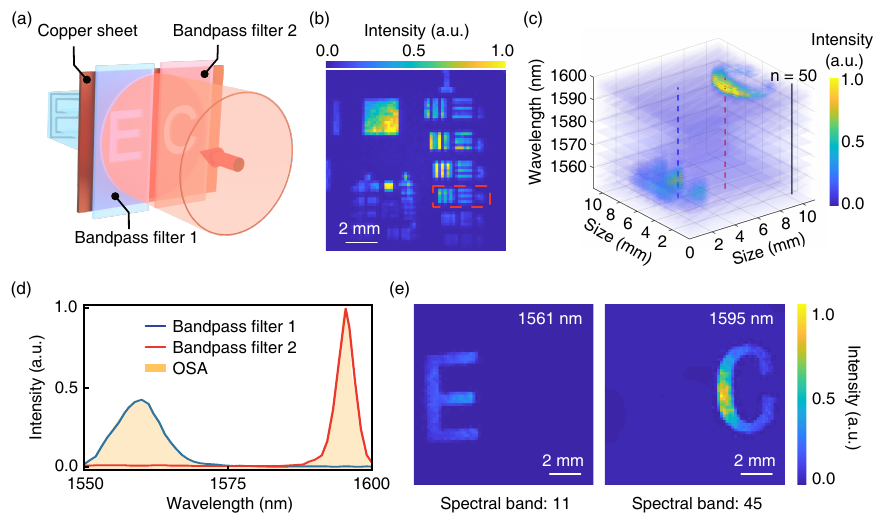}
\caption{Hyperspectral imaging for a static sample. (a) Sample preparation with two types of bandpass filters that cover two letters engraved on a piece of copper sheet. (b) Reconstructed monochromatic image at 1560 nm for a USAF-1951 resolution target. (c) The hyperspectral imaging is measured at a wavelength step of about 1 nm for 50 bands from 1550 to 1600 nm. The full set of spectral images is given in Supplementary Video 1. Dashed lines indicate two representative positions for the reconstructed dataset. (d) Reconstructed spectra at various positions as denoted in (c), which agree well with the spectrum measured by the OSA. (e) Two featured images at spectral bands $n$ = 11 and 45 as denoted in (c).}
\label{fig4}
\end{figure*}

\begin{figure}[t!]
\includegraphics[width=0.85 \columnwidth]{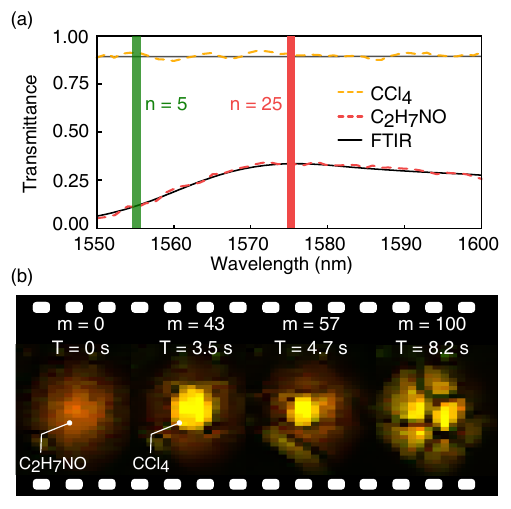}
\caption{Real-time spectral imaging to monitor liquid injection dynamics. (a) Measured transmission spectra of CCl$_{4}$ and C$_{2}$H$_{7}$NO from hyperspectral datasets, which agree well with the FTIR traces. (b) Four selected frames from the recorded video of the dynamic process as shown in Supplementary Video 2. Note that each frame in false color is generated by merging two monochromatic images at spectral channels $n$ = 5 and 25 that are represented in distinct RGB color formats. $m$ denotes the numeration of recorded datacubes.}
\label{fig5}
\end{figure}

Next, we demonstrate the hyperspectral imaging performance by building a static scene. Two types of bandpass filters specified with center wavelengths of 1550 nm and 1600 nm were deployed on a copper sheet engraved with the letters of ``EC" as shown in Fig. \ref{fig4}(a). Note that the spatial resolution was measured using a USAF-1951 resolution target as shown in Fig. \ref{fig4}(b), where the element 4 in group 1 was resolved, corresponding to a spatial resolution of 177 $\mu$m. The spatial resolution is determined by the imaging configuration and the resolvable pixel size of the DMD. In the experiment, an area of 256$\times$256 pixels on the DMD was used to match the 4-fold demagnified image plane, with simultaneous operation of 4$\times$4 pixel binning to enhance the signal-to-noise ratio. The programmable advantage of the DMD allows us to impose flexible control over spatial resolution, field-of-view, and signal-to-noise ratio, which favors versatility to adapt various experimental configurations and application purposes \cite{Edgar2019NP}. Imaging performances at different spatial resolutions and sampling ratios are presented in Supplementary Figures 1 and 2 in Supporting Information, respectively. Figure \ref{fig4}(c) presents the reconstructed hyperspectral data cube containing 50 spectral images with resolvable pixels of 64$\times$64. The full dataset of spectral images is provided in Supplementary Video 1. Subsequently, reconstructed spectra from two marked positions in Fig. \ref{fig4}(c) reveal distinct spectral differentiation of the two bandpass filters as illustrated in Fig. \ref{fig4}(d). Figure \ref{fig4}(e) presents two monochromatic images ($n$ = 11 and 45) from the hyperspectral data cube. Each letter is exclusively visible in its respective spectral channel due to the distinct transmission profile. We note that the reconstruction process of the hyperspectral data cube in the experiment requires only a single encoded temporal waveform per spatially modulated pattern duration, thus significantly alleviating subsequent data storage overhead. Meanwhile, the averaging operation of multiple identical time-domain waveforms is feasible when it comes to improving the signal-to-noise ratio for hyperspectral image reconstruction.

Finally, we investigate real-time hyperspectral imaging to observe liquid injection dynamics. In this scenario, carbon tetrachloride (CCl$_{4}$) was gently injected into ethanolamine (C$_{2}$H$_{7}$NO) in a cuvette. Figure \ref{fig5}(a) presents the measured transmission spectra for two types of liquids, which agree well with the references measured by the FTIR. The chemical differentiation between the CCl$_{4}$ and C$_{2}$H$_{7}$NO was visualized by representing two monochromatic images at spectral channels $n$ = 5 and 25 as green (R=0, G=255, and B=0) and red (R=255, G=0, and B=0) channels, followed by color merging into a single frame \cite{Knez2022SA}. As shown in Fig. \ref{fig5}(b), the C$_{2}$H$_{7}$NO is labeled red color due to its dominant intensity in the red channel, while the CCl$_{4}$ appears yellow owing to comparable intensities across both channels.  Due to the density difference between the two liquids, as CCl$_{4}$ is slowly injected into C$_{2}$H$_{7}$NO, the recorded datacubes ($m$) evolve over time. The yellow region (CCl$_{4}$) in the field of view gradually sinks to the bottom, while the red region (C$_{2}$H$_{7}$NO) is pushed out of view. Eventually, the field of view reveals the shapes of multiple droplets of CCl$_{4}$. To increase the frame rate, compressive sensing strategy is adapted with a sampling ratio of 0.4. Consequently,  we demonstrated real-time monitoring of liquid injection dynamics at a volumetric rate of 12 Hz with resolvable pixels of 32$\times$32 (see Supplementary Video 2). Notably, all spectral channels are simultaneously acquired within a single-frame period, and hence increasing the channel count does not compromise the imaging frame rate.  Furthermore, a lower sampling ratio can be achieved by using more advanced compressive sensing algorithms or deep learning \cite{Huang2022LSA}, which enables a refreshing rate of over 100 Hz. While TV regularization ensures stable reconstructions under constrained sampling, incorporating advanced methods such as Deep Image Prior may further enhance reconstruction fidelity by better preserving image structure and reducing artifacts \cite{Wang2023Photonics}.

\section{Discussions and conclusion}
NIR hyperspectral imaging enables label-free and non-invasive chemical analysis with high-resolution morphologic visualization, emerging as an attractive technique in materials science and medical diagnostics. However, conventional NIR hyperspectral imaging systems rely on costly NIR FPAs combined with inefficient spectral-resolved configurations that significantly increase system complexity and expense \cite{Li2018IPT}. Here, we present a novel single-pixel spectral imaging architecture, leveraging both merits of time-stretch spectroscopy and single-pixel imaging. The involved wavelength-to-time mapping provides an effective solution to circumvent a bandwidth-resolution trade-off and/or low optical throughput, due to the grating-based dispersion \cite{Pian2017NP} or spectral filtering \cite{Welsh2013OE} in the spatial domain. Meanwhile, the single-pixel imaging framework is employed to replace the pricey NIR FPA with a programmable SLM and a single-element detector, allowing flexible optical-field manipulation and improved resistance to noise \cite{Meng2024LSA}. Consequently, the single-pixel detector is used for conducting spatial-spectral multiplexed measurements in the time domain, simultaneously offering high-sensitivity and high-resolution advantages.

We note that multiple improvements are available to surpass the achieved performances. First, the spectral bandwidth and resolution can be further improved to obtain an increased number of spectral channels. To this end, supercontinuum generation can be employed to significantly broaden the laser's spectral output, combined with dispersive elements with larger GVDs and detectors with higher bandwidths, while the laser is adapted to operate at a lower repetition rate. By employing compressed sensing approaches \cite{Kawai2024OL, Lei2017IEEE}, it is possible to overcome the intrinsic trade-off between pulse repetition rate and total GVD in the time-stretch process, thereby enabling the simultaneous achievement of broadband spectral coverage and high spectral resolution. Second, the programmable flexibility of SLMs enables selective interrogation of specific regions with high spatial resolution. Warped sampling for the area of interest can be achieved by the adaptive foveated imaging vision \cite{Phillips2017SA} or the sinusoid structural illumination scheme \cite{Zhang2015NC}. Third, the combination of the time-coding strategy allows us to facilitate synergistic manipulation in the spatio-spectral domain, which could substantially reduce the amount of acquired data while relieving the stringent requirement on the detector's bandwidth \cite{Teng2020AP, Duan2019OL}. Note that in scenarios involving highly scattering or dispersive media, the fidelity and integrity of the time-stretched signal will inevitably be compromised, thereby degrading the reconstruction accuracy of hyperspectral imaging. In such cases, the appropriate incorporation of time-encoding strategies is essential, as their inherent robustness can further enhance acquisition accuracy in complex environments. Fourth, since the encoded temporal waveforms are captured during the elapse of each spatially modulated pattern, the current spectral imaging speed is limited by the frame switching rate of the DMD. Recent advances in MHz-rate pattern projection using spinning cyclic masks \cite{Hahamovich2021NC} or swept aggregate patterns \cite{Kilcullen2022NC} would boost the volumetric rate of the hyperspectral imager to tens of kHz, enabling applications in high-speed dynamics such as biological metabolism monitoring and gas leakage detection. As the frame rate increases, the reduced number of identical temporal waveforms within each spatial modulation pattern limits averaging for SNR improvement, especially under low-light or weak-signal conditions, thereby slightly degrading the fidelity and reconstruction accuracy of hyperspectral imaging.

In summary, we have devised and implemented a single-pixel time-stretch NIR hyperspectral imaging system. The 3D spatial-spectral information is multiplexed into the 1D temporal sequence without complex spectroscopic devices, making it highly attractive for situ applications demanding lightweight and compact size. No additional time is needed to acquire the spectral information compared to conventional single-pixel imaging, which preserves the inherent simplicity of single-pixel detection while achieving rapid frame rates and high-throughput operation. Furthermore, the presented single-pixel time-stretch hyperspectral imaging modality can be extended to mid-/far-infrared or terahertz wavelengths through frequency upconversion detection, addressing the critical lack of high-performance detectors and high-fidelity modulators in these spectral regions \cite{Wang2023NC, Sun2025LPR}. Additionally, integration with time-correlated coincidence counting would enable efficient hyperspectral imaging at the single-photon level \cite{Sun2024LPR}. When further combined with time-of-flight techniques, this architecture holds potential for higher-dimensional imaging (i.e., $x, y, z, \lambda$), thereby opening up novel possibilities in photon-starved scenarios.\\

\section*{Acknowledgements}
This work was funded by Shanghai Pilot Program for Basic Research (TQ20220104); National Natural Science Foundation of China (62175064, 62235019, 62035005); Innovation Program for Quantum Science and Technology (2023ZD0301000); Shanghai Municipal Science and Technology Major Project (2019SHZDZX01); Natural Science Foundation of Chongqing (CSTB2023NSCQ-JQX0011, CSTB2022TIAD-DEX0036, CSTB2025NSCQ-GPX0443);  Fundamental Research Funds for the Central Universities (YBNLTS2025-009); China Post doctoral Science Foundation (2024M760918); Postdoctoral Fellowship Program and China Postdoctoral Science Foundation (GZC20250545, 2024M760918, 2025T180224).

\section*{Conflict of Interest}
The authors declare no conflict of interests.

\section*{Supporting Information}
Supporting Information is available from the Wiley Online Library or from the author.

\section*{Data Availability Statement}
The data that support the findings of this study are available from the corresponding author upon reasonable request.

\section*{Keywords}
spectral imager, near-infrared hyperspectral imaging, time-stretch spectroscopy, single-pixel imaging


\begin{thebibliography}{100}

\bibitem{Khan2018IEEE} M. J. Khan, H. S. Khan, A. Yousaf, K. Khurshid and A. Abbas, ``Modern Trends in Hyperspectral Image Analysis: A Review," IEEE Access \textbf{6}, 14118-14129 (2018).

\bibitem{Ozaki2021Book} Y. Ozaki, C. Huck, S. Tsuchikawa and S. B. Engelsen, \textit{Near-Infrared Spectroscopy: Theory, Spectral Analysis, Instrumentation, and Applications} (Springer, Singapore, 2021).

\bibitem{Wang2024NP} F. Wang, Y. Zhong, O. Bruns, Y. Liang and H. Dai, ``In vivo NIR-II fluorescence imaging for biology and medicine," Nat. Photonics \textbf{18}, 535-547 (2024).

\bibitem{Rodrigues2022ABC} E. M. Rodrigues and E. Hemmer, ``Trends in hyperspectral imaging: from environmental and health sensing to structure-property and nano-bio interaction studies," Anal. Bioanal. Chem. \textbf{414}, 4269-4279 (2022).

\bibitem{Stegemann2025AS} J. Stegemann, F. Gr$\ddot{\text{o}}$niger, K. Neutsch, H. Li, B. S. Flavel, J. T. Metternich, L. Erpenbeck, P. B. Petersen, P. N. Hedde and S. Kruss, ``High-Speed Hyperspectral Imaging for Near Infrared Fluorescence and Environmental Monitoring," Adv. Sci. \textbf{12}, 2415238 (2025). 

\bibitem{Liu2017TFST} Y. Liu, h. Pu and D. Sun, ``Hyperspectral imaging technique for evaluating food quality and safety during various processes: a review of recent applications," Trends Food Sci. Technol. \textbf{69}, 25-35 (2017).

\bibitem{Li2024NC} D. Li, J. Wu, J. Zhao, H. Xu and L. Bian, ``SpectraTrack: megapixel, hundred-fps, and thousand-channel hyperspectral imaging," Nat. Commun. \textbf{15}, 9459 (2024).

\bibitem{Arce2014ISPM} G. R. Arce, D. J. Brady, L. Carin, H. Arguello and D. S. Kittle, ``Compressive coded aperture spectral imaging: An introduction," IEEE Signal Process. Mag. \textbf{31}, 105-115 (2014).

\bibitem{Roxbury2015SR} D. Roxbury, P. V. Jena, R. M. Williams, B. Enyedi, P. Niethammer, S. Marcet, M. Verhaegen, S. Blais-Ouellette and D. A. Heller, ``Hyperspectral Microscopy of Near-Infrared Fluorescence Enables 17-Chirality Carbon Nanotube Imaging," Sci. Rep. \textbf{5}, 14167 (2015).

\bibitem{Kaariainen2021OL} T. K$\ddot{\text{a}}$$\ddot{\text{a}}$ri$\ddot{\text{a}}$inen and Timo D$\ddot{\text{o}}$nsberg, ``Active hyperspectral imager using a tunable supercontinuum light source based on a MEMS Fabry–Perot interferometer," Opt. Lett. \textbf{46}, 5533-5536 (2021).

\bibitem{Li2018IPT} X. Li, H. Gong, J. Fang, X. shao, H. Tang, S. Huang, T. Li and Z. Huang, ``The development of InGaAs short wavelength infrared focal plane arrays with high performance," Infrared Phys. Technol. \textbf{80}, 112-119 (2017).

\bibitem{Dorling2013TB} K. M. Dorling and M. J. Baker, ``Rapid FTIR chemical imaging: highlighting FPA detectors," Trends Biotechnol. \textbf{31}, 437-438 (2013).

\bibitem{Wang2019NC} Z. Wang, S. Yi, A. Chen, M. Zhou, T. S. Luk, A. James, j. Nogan, W. Ross, G. Joe, A. Shahsafi, K. X. Wang, M. A. Kats and Z. Yu, ``Single-shot on-chip spectral sensors based on photonic crystal slabs," Nat. Commun. \textbf{10}, 1020 (2019).

\bibitem{Bian2024Nature} L. Bian, Z. Wang, Y. Zhang, L. Li, Y. Zhang, C. Yang, W. Fang, J. Zhao, C. Zhu, Q. Meng, X. Peng and J. Zhang, ``A broadband hyperspectral image sensor with high spatio-temporal resolution," Nature \textbf{635}, 73-81 (2024).

\bibitem{Tao2021OE} C. Tao, H. Zhu, X. Wang, S. Zheng, Q. Xie, C. Wang, R. Wu and Z. Zheng, ``Compressive single-pixel hyperspectral imaging using RGB sensors," Opt. Express \textbf{29}, 11207-11220 (2021).

\bibitem{Edgar2019NP} M. P. Edgar, G. M. Gibson and M. J. Padgett, ``Principles and prospects for single-pixel imaging," Nat. Photonics \textbf{13}, 13-20 (2019).

\bibitem{Zhang2018Optica} Z. Zhang, S. Liu, J. Peng, M. Yao, G. Zheng and J. Zhong, ``Simultaneous spatial, spectral, and 3D compressive imaging via efficient Fourier single-pixel measurements," Optica \textbf{5}, 315-319 (2018).

\bibitem{Wu2021NC} D. Wu, J. Luo, G. Huang, Y. Feng, X. Feng, R. Zhang, Y. Shen and Z. Li, ``Imaging biological tissue with high-throughput single-pixel compressive holography," Nat. Commun. \textbf{12}, 4712 (2021).

\bibitem{Sun2016NC} M. Sun, M. P. Edgar, G. M. Gibson, B. Sun, N. Radwell, R. Lamb and M. J. Padgett, ``Single-pixel three-dimensional imaging with time-based depth resolution," Nat. Commun. \textbf{7}, 12010 (2016).

\bibitem{Pian2017NP} Q. Pian, R. Yao, N. Sinsuebphon and X. Intes, ``Compressive hyperspectral time-resolved wide-field fluorescence lifetime imaging," Nat. Photonics \textbf{11}, 411-414 (2017).

\bibitem{Uguen2024APL} L. Uguen, R. Piedevache, G. Russias, S. Helmer, D. Tregoat and S. Perrin, ``Single-pixel-based hyperspectral microscopy," Appl. Phys. Lett. \textbf{125}, 071108 (2024).

\bibitem{Welsh2013OE} S. S. Welsh, M. P. Edgar, R. Bowman, P. Jonathan, B. Sun and M. J. Padgett, ``Fast full-color computational imaging with single-pixel detectors," Opt. Express \textbf{21}, 23068-23074 (2013).

\bibitem{Gattinger2019OE} P. Gattinger, J. Kilgus, I. Zorin, G. Langer, R. Nikzad-Langerodi, C. Rankl, M. Gr$\ddot{\text{o}}$schl and M. Brandstetter, ``Broadband near-infrared hyperspectral single pixel imaging for chemical characterization," Opt. Express \textbf{27}, 12666-12672 (2019).

\bibitem{Ebner2023SR} A. Ebner, P. Gattinger, I. Zorin, L. Krainer, C. Rankl and M. Brandstetter, ``Diffraction-limited hyperspectral mid-infrared single-pixel microscopy," Sci. Rep. \textbf{13}, 281 (2023).

\bibitem{Xu2024NC} Y. Xu, L. Lu, V. Saragadam and K. F. Kelly, ``A compressive hyperspectral video imaging system using a single-pixel detector," Nat Commun. \textbf{15}, 1456 (2024).

\bibitem{Sun2025LPR} B. Sun, K. Huang, H. Ma, J. Fang, T. Zheng, R. Qin, Y. Chu, H. Guo, Y. Liang and H. Zeng, ``Mid-infrared single photon compressive spectroscopy," Laser Photonics Rev. \textbf{19}, 2401099 (2025).

\bibitem{Bian2016SR} L. Bian, J. Suo, G. Situ, Z. Li, J. Fan, F. Chen and Q. Dai, ``Multispectral imaging using a single bucket detector," Sci. Rep. \textbf{6}, 24752 (2016).

\bibitem{Jin2017SR} S. Jin, W. Hui, Y. Wang, K. Huang, Q. Shi, C. Ying, D. Liu, Q. Ye, W. Zhou and J. Tian, ``Hyperspectral imaging using the single-pixel Fourier transform technique," Sci. Rep. \textbf{7}, 45209 (2017).

\bibitem{Meng2024LSA} H. Meng, Y. Gao, X. Wang, X. Li, L. Wang, X. Zhao and B. Sun, ``Quantum dot-enabled infrared hyperspectral imaging with single-pixel detection," Light Sci. Appl. \textbf{13}, 121 (2024).

\bibitem{Xiong2023LSA} J. Xiong, Z. Zhang, Z. Li, P. Zheng, J. Li, X. Zhang, Z. Gao, Z. Wei, G. Zheng, S. Wang and H. Liu, ``Perovskite single-pixel detector for dual-color metasurface imaging recognition in complex environment," Light Sci. Appl. \textbf{12}, 286 (2023).

\bibitem{Solli2008NP} D. R. Solli, J. Chou and B. Jalali, ``Amplified wavelength-time transformation for real-time spectroscopy," Nat. Photonics \textbf{2}, 48-51 (2008).

\bibitem{Goda2013NP} K. Goda and B. Jalali, ``Dispersive Fourier transformation for fast continuous single-shot measurements," Nat. Photonics \textbf{7}, 102-112 (2013).

\bibitem{Hashimoto2023LSA} K. Hashimoto, T. Nakamura, T. Kageyama, V. R. Badarla, H. Shimada, R. Horisaki and T. Ideguchi, ``Upconversion time-stretch infrared spectroscopy," Light Sci. Appl. \textbf{12}, 48 (2023).

\bibitem{Sun2024LPR} B. Sun, K. Huang, H. Ma, J. Fang, T. Zheng, Y. Chu, H. Guo, Y. Liang, E Wu, M. Yan and H. Zeng, ``Single-photon time-stretch infrared spectroscopy," Laser Photonics Rev. \textbf{18} 2301272 (2024).

\bibitem{Zhang2025NRMP} Y. Zhang, C. Tao, S. Luo, K. Y. Lau, J. Zheng, L. Huang, A. Zhang, L. Sheng, Q. Ling, Z. Guan, Y. Cui, D. Chen, J. Qiu, S. K. Turitsyn and Z. Sun, ``Ultra-fast optical time-domain transformation techniques," Nat. Rev. Methods Primers \textbf{5}, 11 (2025).

\bibitem{Cheng2016APR} C. Lei, B. Guo, Z. Cheng and K. Goda, ``Optical time-stretch imaging: Principles and applications," Appl. Phys. Rev. \textbf{3}, 011102 (2016).

\bibitem{Goda2009Nature} K. Goda, K. K. Tsia and B. Jalali, ``Serial time-encoded amplified imaging for real-time observation of fast dynamic phenomena," Nature \textbf{458}, 1145-1149 (2009).

\bibitem{Lau2016LOAC} A. K. S. Lau, H. C. Shum, K. K. Y. Wonga and K. K. Tsia, ``Optofluidic time-stretch imaging - an emerging tool for high-throughput imaging flow cytometry," Lab Chip \textbf{16}, 1743-1756 (2016).

\bibitem{Teng2020AP} J. Teng, Q. Guo, M. Chen, S. Yang and H. Chen, ``Time-encoded single-pixel 3D imaging," APL Photonics \textbf{5}, 020801 (2020).

\bibitem{Duan2019OL} Y. Duan, X. Dong, N. Yang, C. Zhang, K. K. Y. Wong and X. Zhang, ``Temporally structured illumination for ultrafast time-stretch microscopy," Opt. Lett. \textbf{44}, 4634-4637 (2019).

\bibitem{Thomas2018OE} T. Gerrits, D. J. Lum, V. Verma, J. Howell, R. P. Mirin and S. W. Nam, ``Short-wave infrared compressive imaging of single photons," Opt. Express \textbf{26}, 15519-15527 (2018).

\bibitem{Jiang2022PR} W. Jiang, Y. Yin, J. Jiao, X. Zhao and B. Sun, ``2,000,000 fps 2D and 3D imaging of periodic or reproducible scenes with single-pixel detectors," Photonics Res. \textbf{10}, 2157 (2022).

\bibitem{Duarte2008IEEE} M. F. Duarte, M. A. Davenport, D. Takhar, J. N. Laska, T. Sun, K. F. Kelly and R. G. Baraniuk, ``Single-pixel imaging via compressive sampling," IEEE Signal Process. Mag. \textbf{25}, 83-91 (2008).

\bibitem{Shoshin2025LPR} M. Shoshin, T. Kageyama, T. Nakamura, K. Hashimoto and T. Ideguchi, ``Mid-infrared frequency-swept laser at 50 MScans/s via frequency down-conversion of time-stretched near-infrared pulses," Laser Photonics Rev. \textbf{19}, 2500008 (2025).

\bibitem{Wang2023NC} Y. Wang, K. Huang, J. Fang, M. Yan, E Wu and H. Zeng, ``Mid-infrared single-pixel imaging at the single-photon level," Nat. Commun. \textbf{14}, 1073 (2023).

\bibitem{Cai2024SA} Y. Cai, Y. Chen, K. Dorfman, X. Xin, X. Wang, K. Huang and E Wu, ``Mid-infrared single-photon upconversion spectroscopy enabled by nonlocal wavelength-to-time mapping," Sci. Adv. \textbf{10}, eadl3503 (2024).

\bibitem{Knez2022SA} D. Knez, B. W. Toulson, A. Chen, M. H. Ettenberg, H. Nguyen, E. O. Potma and D. A. Fishman, ``Spectral imaging at high definition and high speed in the mid-infrared," Sci. Adv. \textbf{8}, eade4247 (2022).

\bibitem{Huang2022LSA} L. Huang, R. Luo, X. Liu and X. Hao, ``Spectral imaging with deep learning," Light Sci. Appl. \textbf{11}, 61 (2022).

\bibitem{Wang2023Photonics} C. Wang, H. Li, S. Bie, R. Lv and X. Chen, ``Single-Pixel Hyperspectral Imaging via an Untrained Convolutional Neural Network," Photonics \textbf{10}, 224 (2023).

\bibitem{Kawai2024OL} A. Kawai, R. Horisaki and T. Ideguchi, ``Compressive time-stretch spectroscopy with pulse-by-pulse intensity modulation," Opt. Lett. \textbf{49}, 3468-3471 (2024).

\bibitem{Lei2017IEEE} C. Lei, Y. Wu, A. C. Sankaranarayanan, S.-M. Chang, B. Guo, N. Sasaki, H. Kobayashi, C.-W. Sun, Y. Ozeki and Keisuke Goda, ``GHz optical time-stretch microscopy by compressive sensing," IEEE Photonics J. \textbf{9}, 1-8 (2017).

\bibitem{Phillips2017SA} D. B. Phillips, M. J. Sun, J. M. Taylor, M. P. Edgar, S. M. Barnett, G. M. Gibson and M. J. Padgett, ``Adaptive foveated single-pixel imaging with dynamic supersampling," Sci. Adv. \textbf{3}, e1601782 (2017).

\bibitem{Zhang2015NC} Z. Zhang, X. Ma and J. Zhong, ``Single-pixel imaging by means of Fourier spectrum acquisition," Nat. Commun. \textbf{6}, 6225 (2015).

\bibitem{Hahamovich2021NC} E. Hahamovich, S. Monin, Y. Hazan and A. Rosenthal, ``Single pixel imaging at megahertz switching rates via cyclic Hadamard masks," Nat. Commun. \textbf{12}, 4516 (2021).

\bibitem{Kilcullen2022NC} P. Kilcullen, T. Ozaki and J. Liang, ``Compressed ultrahigh-speed single-pixel imaging by swept aggregate patterns," Nat. Commun. \textbf{13}, 7879 (2022).


\end{thebibliography}
\end{document}